\documentclass{article}
\usepackage{times} 
\include{psfig}
\usepackage{epsfig}

\begin{document}

\title{Socioeconomic Development and Stability: \\
A Complex Network Blueprint}

\author{Luciano da Fontoura Costa, \\
Instituto de F\'{\i}sica de S\~ao Carlos. \\ 
Universidade de S\~ao Paulo, S\~{a}o Carlos, \\
SP, Caixa Postal 369, 13560-970, \\
Fone +55 16 3373 9858, FAX +55 1633 71 3616, \\
Brazil, luciano@if.sc.usp.br}

\date{21st March 2005}

\maketitle

{\bf ABSTRACT:} Complex networks provide flexible and powerful
resources for characterizing, modeling and simulating a wide range of
real world complex systems.  The current work discusses how such a
versatile new area can be applied in order to aid economical
development and stability at several scales and contexts.  The
following activities are involved: (a) compilation of several types of
data related to socioeconomic development, including several types of
transportation systems, availability of human and natural resources,
communication and energy networks, climate and geographical features,
as well as endemic diseases, to name but a few; (b) representation of
such data in terms of multilayer interacting complex networks
(cond-mat/0406369) registered geographically; (c) application of
traditional and new methods for complex networks characterization and
analysis.  Such an approach allows the identification of bottlenecks
and deficits/surpluses, simulation of system development under varying
constraints and perturbations, as well as the application of
optimization methods in order to help identify the most effective
strategies leading to social and economic wealth and stability.  In
addition to its practical implications, such an approach also
emphasizes several issues of substantial theoretical interest,
including the integration of networks of different natures, the
interplay between the dynamics of topological and node state
evolution, the effects of geographical constraints, community finding,
as well as the interesting problem of how to optimize such systems
with respect to network topology and dynamics in order to achieve
specific objectives expressed by merit figures.  The discussed
methodology is particularly interesting for applications in developing
countries because of the greater potential for economic evolution in
such nations.  This manuscript also includes a brief review of complex
networks approaches to socioeconomics modeling.

\pagebreak

\begin{quotation}
'... it is hard to turn ideas into wealth in the absence of social 
connectedness, which in the age of the Internet still requires 
something more than bandwidth and high-speed connectivity.' \\
\hspace{2cm} \emph{(F. Fukuyama, The Great Disruption)}
\end{quotation}

\section{Introduction}

The main activity in science is the art of model building.  Good
models require sound representation of the phenomenon of interest in
terms of a reduced set of most relevant variables and parameters.
Because all models are necessarily incomplete, it is important to
obtain particularly effective representations, which should be able to
account for the relevant variables under constraints imposed by the
chosen parameters.  By being naturally oriented to representing
connections and relationships, graph theory stands out as particularly
general and suitable for model representations.  Indeed, almost all
discrete data structures can be understood as a particular instance of
graphs.  As the relationships between variables and parameters
typically change with time, it is also necessary that the adopted
representations be capable of expressing dynamical changes in both
network topology as well as the dynamics undergone by the states of
the nodes.  With this respect, the seminal works by Flory~\cite{Flory},
Rapoport~\cite{Rapoport}, Erd\"os and R\'enyi~\cite{Erdos_Renyi} on
random networks have been substantially expanded in the recent years
mainly through the consideration of principles from statistical
physics and dynamic systems, especially regarding the characterization
and modeling of scale free network models~\cite{Albert_Barab:2002}.

A direct indication of the impressive success of complex network
research is the large and ever increasing related scientific
production, accounting for about 1 or 2 new \emph{arxiv} manuscripts
per day (\emph{http://arxiv.org/archive/}).  Indeed, the catalisis
of scientific investigation, interaction and dissemination allowed by
the WWW and internet resources as \emph{arxiv} provides further
evidence of the impact of immediate wide dissemination of data,
results and knowledge.  One particularly interesting portion of the
developments in complex network has been aimed at applying concepts,
measurements and models in order to mimic and predict the behavior of
real complex systems including the internet (e.g.~\cite{Faloutsos,
peering, Barth_internet, Barth_Gondran}), WWW (e.g.~\cite{Bara_WWW,
Kahng_WWW}), protein and metabolic interactions
(e.g.~\cite{Jeong_protein, Reka_protein}), transportation systems
(e.g.~\cite{Bagler, Parongama, Guimera_airport}), opinion formation
(e.g.~\cite{Elgazzar, forsta, Stauffer_Hohnisch, L_sznaj:2005}),
epydemiology (e.g.~\cite{Masuda_diseases, Szendroi_diseases}), among
many other relevant issues.  Because of its flexibility and power for
developing good integrative models of complex systems, it becomes
particularly interesting to consider the systematic application of
this new area as means to represent, model, predict and optimize
the behavior of socioeconomic systems.

Socioeconomics environments rank among the most complex systems known
to humanity, involving a whole range of entities and relationships.
By selecting a set of reference entities (e.g. institutions or
cities), it is possible to obtain a series of integrated complex
networks~\cite{Solomon} registered by geography, each representing
specific types of relationships.  Several concepts, measurements and
modeling approaches derived from the area of complex network research
can then be applied in order to characterize several topological and
dynamical features of such systems, allowing the prediction of the
response of the system to modifications, the identification of
promising development strategies, the enhancement of the stability of
the whole system, optimization in order to achieve specific
objectives, as well as its resilience to disruptions (e.g. epidemics
and catastrophes), to name but a few possibilities.

This text describes a possible blueprint for such investigations,
which are relevant not only for their immediate implications and
potential for optimizing commonwealth, but also for poising several
interesting theoretical issues arising from the integration of
networks and optimization of topology-dynamics under pre-specified
optimality indices.  It should be emphasized at the outset that it is
the broad integrative perspective of considering several types of
evolving geographically registered geographical networks, with
emphasis on applications to third-world socioeconomics, which
provides the main motivation for the present work.  It should also be
noted that any result arising from such a type of simulations need to
be treated with the greatest caution because of the several non-linear
and unpredictable effects which are known to operate in
socioeconomics.  However, it is expected that the systematic and
integrative use of concepts and methods from complex network research,
itself a very dynamic research area, can provide valuable means for
optimizing socioeconomic development and stability.

This article starts by briefly reviewing some of the main complex
network approaches to socioeconomics and follows by discussing the
representation of the modeled systems in terms of multi-layer,
geographically registered networks, as well as simulation and
optimization possibilities.

\section{A Brief Review of Complex Network Approaches to \\
Socioeconomics and Related Issues}

Previous developments involving complex network approaches to
socioeconomics reported in the literature are briefly reviewed in the
following.  One of the first socioeconomics related applications of
complex networks was reported by Guardiola et
al.~\cite{Guimera_trusts}, where the web of trust scheme between users
of PGP was considered as a model for trust networks, leading to good
resilience to intentional attack.  The use of small-world models for
socioeconomic systems was investigated by Elgazzar~\cite{Elgazzar}
through simulation of the Sznajd dynamics~\cite{Sousa, forsta} on
small-world networks.  The issue o geographical embedding of
networks~\cite{Rozenfeld, Alon, Andersson} as a constraint to
U.S. Internet infrastructure and its implications for economy and
politics was considered in~\cite{Gorman}.  Spatial constraints have
also been found to affect networking and internet
architectures~\cite{Gray}, in the sense that computing resources would
tend to be placed as close as possible to the source of data in order
to avoid expensive network traffic.  The network topology of the
Austrian Interbank market was investigated by Boss and
collaborators~\cite{Thurner}, indicating that the contract size
distribution follow an extensive power law.  The issue of grid
computing has been addressed in terms of resource allocation and
regulation~\cite{Buyya, Lai} as well as by adopting complex network
interconnections~\cite{Costa_grid}. Bonanno et al.~\cite{Bonanno}
investigated the possibility to extract meaningful economic
information from portfolio of stocks and its implications for
comparison of the topological properties of networks.  A model of
wealth dynamics and transactions among economic agents by considering
different network connectivities was investigated by Garlaschelli and
Loffredo~\cite{Garlaschelli}.  The effect of information cascades over
economic recessions in the U.S., assuming a random network model was
described by Cook~\cite{Cook}, while studies of cascades of failures
as a consequence of attacks have been covered in~\cite{Hayashi}.  The
possibility of representing economic variables of different nature in
terms of multiple interconnected networks, called \emph{Solomon
networks}, have been considered for Ising simulations of the dynamics
of economic systems~\cite{Solomon}.  Bipartite network models have
been used to simulate relationship between countries and currencies in
world exchange arrangements web~\cite{Li}, considering assortativity
aspects.  The stability in supply/production networks has been
investigated by Helbing~\cite{Helbing} who analysed, by considering
different network topologies, how networks of damped oscillators tend
to be subject to increasing oscillations.  Topological investigations
of networks defined by traders exchanging goods have been considered
by Reichardt and Bornhold~\cite{Reichardt}, who analysed the 2004
pre-Christmas season and identified high modularity.  A review of
econophysics has been presented by Di~\cite{Di}, and a review of
quantitative modelling of financial markets has been reported by
Farmer and Lo~\cite{Farmer}.

\section{Representation}

One of the first important decisions while modeling a socioeconomic
system is to define and represent in a careful way its most
representative components.  Such components can be divided into two
categories: \emph{states} and \emph{relationships}. In socioeconomics
networks, local states correspond to properties of the main considered
sites or places (e.g. cities, institutions, etc.).  Examples of states
include but are not limited to:

\begin{trivlist}
  \item (a) \emph{Human resources}: The involved individuals, possibly
  subdivided into workers, consumers, experts, etc.  

  \item (b) \emph{Natural resources}: The existing (or prospected)
  energy sources, organic and inorganic assets, rivers and lakes,
  climatic features, etc.

  \item (c) \emph{Storage capabilities}: The local potential for
  storing raw and processed materials.

  \item (d) \emph{Industrial resources}: The facilities available
  locally which can be used to process raw materials as well as high
  technological means for obtaining more sophisticate goods.

  \item (e) \emph{Financial resources}: May include the bank and
  finance systems which can be found in the locality.

  \item (f) \emph{Endemic diseases}: The epidemics and pathologies
  which continuous or periodically affect humans and animals in the
  region.

  \item (g) \emph{Cultural and social features}: The cultural
  traditions and social features and values.

  \item (g) \emph{Scientific and technological assets}: The level of
  scientific and technological development at each locality.

\end{trivlist}

Note that several of such states are not straightforward to be
quantified, implying some degree of arbitrariness.

\begin{figure}
  \begin{center}
    \includegraphics[scale=.7]{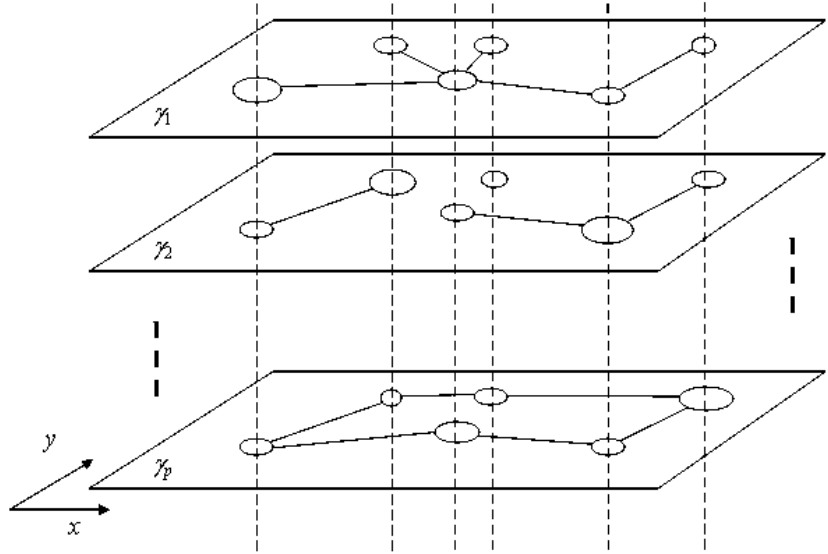}  

    \caption{Socioeconomics complex network models may involve
    several geographical networks characterized by the fact that the
    nodes have definite spatial positions.  Each network $\gamma_i$
    represents one of the considered types of relationship.  Note that
    the registration between these networks is accomplished through
    the spatial congruence of the nodes position. The interactions
    between the several layers are not represented in this
    diagram.}~\label{fig:exnet}

  \end{center}
\end{figure}

Global states are typically maps of local states into overall
properties of the modeled system, such as overall production, debt or
surplus, total birth/death rates, etc.  The relationships between the
localities follow naturally, defining a complex network for each
considered type of interaction~\cite{Solomon}.  A particularly
interesting possibility is to integrate such networks geographically,
i.e. each site is represented as a node with geographical position and
several types of edges are defined between such common nodes, yielding
an integration of several geographical networks in a way that reminds
the topographical connections between cortical
layers~\cite{Luc_topo}. Figure~\ref{fig:exnet} illustrates a simple
hypothetical socioeconomic model involving six cities and $p$
relationship networks.

Examples of relationships relevant for socioeconomic systems are
listed in the following:

\begin{trivlist}
  \item (A) \emph{Transportation}: Essential for economic integration,
  allowing human, raw materials, and processed goods displacement.
  Each type of transportation (e.g. railway, motorways, airways) can
  be represented as a specific complex network, facilitating the
  analysis of the complementarity between such resources.

  \item (B) \emph{Energy distribution}: The existing network allowing
  access to several types of energy (e.g. electrical and gas).

  \item (C) \emph{Communications}: The interconnections allowing
  information exchange between the considered localities, possibly
  subdivided into networks for cell and fixed telephony, satellite,
  optic fibers, etc.

  \item (D) \emph{Financial and political trusts and alliances}:
  Corresponding to the network of government and private financial
  agencies.

  \item (E) \emph{Borrow/Loan relationships}: The directed network
  defined by borrowing and debts between the involved localities.

  \item (F) \emph{Cultural and social links}: Networks
  established by common share of common beliefs and traditions.

  \item (G) \emph{Distributed computing}: Including wide range
  distributed computing facilities, such as grid computing.

\end{trivlist}

Observe that most such networks are typically dynamical (in the sense
that their topology will vary with time) and weighted.  Flow or
resources conservation may eventually be observed.  Except for
communications networks, all the above networks are typically
represented as digraphs (i.e. involving oriented edges).

\section{Characterization and Simulation}

Once a part of a socioeconomic system has been represented in terms
of geographically integrated networks, a series of measurements can be
used to characterize and analyse their topology.  The choice of
measurements should be performed with basis on the specific issues of
interest. For instance, in the case of transportation systems,
statistics of shortest paths are of special relevance.  The reader is
reported to a recent survey~\cite{Luc_rev} for a comprehensive review
of complex network measurements.  Of special interest are the specific
demands implied for the topological characterization of the
geographically registered coexisting networks, motivating new
measurements capable of expressing the topological interactions
between the several layers (e.g.~\cite{Luc_topo}).  Another
particularly relevant issue is the identification of communities in
the integrated networks.  A particularly interesting possibility is to
consider the identification of well-defined communities in one of the
layers as a subsidy for identification of communities in the remainder
layers, as well as the analysis of overlaps and divergences between
such clusters.  Perspectives of special interest also include the
development of models describing the topological evolution of the
networks at each layer and as a whole.

The simulation of the dynamics of the network states can be performed
by assuming several methods including spin dynamics
(e.g.~\cite{Solomon, Bose}), cellular automata, and systems of coupled
differential equations.  Global feedback can also be considered in
such formulations~\cite{Solomon}.  Important related aspects involve
the synchronization of events in the networks as well as the
appearance of instabilities and oscillations~\cite{Helbing}.  A
particularly interesting possibility is to link the dynamics of the
network individual states with the dynamics of topological changes in
the network structure (e.g.~\cite{L_sznaj:2005}).

While the simulation of the dynamical evolution of the network states
can lead to valuable insights about socioeconomic development and
stability, it is also interesting to consider the \emph{optimization}
of the network architecture and constraints in order to achieve
specific goals.  Provided merit figures are clearly established in
terms of properties of the modeled system, a series of optimization
approaches ranging from linear programming to genetic algorithms can
be applied in order to identify improvements to the topology of the
network or its dynamical evolution.  For instance, once a government
have decided to explore a recently discovered source of raw material
or to find the best way to protect indigenous fauna and flora,
optimization of the topology of the network under constrains imposed
by its states can be performed in order to evaluate possible
development strategies.  Of course, such approaches are inherently
limited by the suboptimality (i.e. convergence to local minima)
characterizing most non-linear optimization methods.

\section{Concluding Remarks}

This article has discussed how complex network concepts and methods
can be extensively used to model socioeconomic systems.  After
reviewing briefly the main related literature, a blueprint has been
proposed and discussed.  Such a model involves multilayer
networks~\cite{Solomon} whose nodes are registered in terms of the
geographical positions of the entities which they
represent~\cite{Luc_topo}.  Several interesting issues are motivated
regarding the characterization and simulation of such networks.  As
far as the analysis of the topological features of the multiple
networks are concerned, adaptations of traditional measurements and
new features capable of taking into account and quantifying the
interconnections between different layers are of special interest.
Two mains possibilities are defined regarding the dynamics of state
evolution: (i) the simulation of the dynamics on static diverse
topologies; and (ii) the simulation of state dynamics on networks
whose topology undergoes dynamic evolution.

Several are the difficulties constraining such investigations.  To
begin with, it is important to gather reliable, uptodated and
representative data related to each socioeconomic features considered
in the model.  A particularly challenging issue concerns the
definition of the merit figures used for the optimization of the
topology and weights of the network, which often implies political,
ethical and/or arbitrary nature (e.g. should transgenic products or
abortion be allowed?).  Also, the high complexity of the involved
systems severely constraints the time window for predictions, implying
that the greatest caution should be taken when analysing
characterization and simulation results.  Despite such difficulties,
it should still be possible to adopt a progressive approach starting
with only a few layers and gradually increasing complexity.  An
example of a particularly feasible and interesting starting point is
to investigate the efficiency of coexisting transportation networks
(e.g. railways, motorways, airport systems) while trying to identify
how such systems can be improved (e.g. minimize the average shortest
path) by small topological modifications (e.g. the inclusion of a new
railway link).  Such models can be easily upgraded by including the
availability of natural and human resources, consequently defining
interesting problems of optimizing flow and production-consuming
interactions.

The author is currently conducting related efforts considering the
Brazilian economy and would highly appreciate to receive comments and
suggestions and to consider collaborations in any theoretical or
practical related aspects.
 
\bibliographystyle{unsrt}
\bibliography{proj}

\end{document}